\documentclass{ws-procs975x65}
\usepackage{graphicx}
\usepackage{hyperref}
\usepackage{cite}
\usepackage{xspace}

\newcommand{\e}[1]{\mathrm{e}^{#1}}

\def\GRG{Gen.\ Relat.\ Gravit.} 

\def\CQG{Classical Quant.\ Grav.}

\def\JMP{J.\ Math.\ Phys.}

\def\PRD{Phys.\ Rev.\ D}

\def\PR{Phys.\ Rev.}

\def\PRSLA{Proc.\ Roy.\ Soc.\ London A}%

\def\RMP{Rev.\ Mod.\ Phys.}

\begin{document}
\title{\uppercase{Source Integrals for
Multipole Moments in Static Spacetimes}}

\author{\uppercase{Norman G{\"u}rlebeck}}

\address{ZARM, University of Bremen, Am Fallturm, 28359 Bremen,
Germany, EU\\
Institute of Theoretical Physics, Charles University, V
Hole\v{s}ovi\v{c}k\'ach 2, 180 00 Praha, Czech Republic, EU\\
E-mail: norman.guerlebeck@zarm.uni-bremen.de}


\begin{abstract}
We derive source integrals for multipole moments that describe the behavior of
static and axially symmetric spacetimes close to spatial infinity. We
assume that the matter distribution is isolated. We outline also some
applications of these source integrals of the asymptotic multipole moments.
\end{abstract}

\keywords{Multipole moments, static spacetimes, source integrals}

\bodymatter

\section{Introduction}

The multipole moments of a Newtonian mass distribution can be determined in
different ways. First, we can expand the gravitational potential far away from
the source. The coefficients of the resulting series are then identified with
the multipole moments. Secondly, the moments of the mass density $\mu$, i.e.
integrals of products of $\mu$ and certain polynomials, equal the multipole
moments. For instance, we have for axially symmetric distributions and the
multipole moments $M_i$ in polar coordinates $(r,\theta,\varphi)$:
\begin{align*}
   U&=-\sum\limits_{i=0}^{\infty} \frac{M_i}{r^{i+1}}P_i(\cos\theta),\quad M_i=\int\limits_V\mu r^i P_i(\cos\theta){\mathrm d}V,
\end{align*}
where $P_i$ denote the Legendre polynomials of the first kind and $V$ is the
support of $\mu$. Geometric units are chosen such that $G=c=1$. That these
two definitions are equivalent is not trivial but can be proved using the
Poisson integral. In general relativity, both types of definitions exist but
they are in general not equivalent.
The first type of definition involves only the gravitational field close to
space-like or null-like infinity and it yields \emph{asymptotic multipole
moments} (or field multipole moments\cite{Ashtekar_2004}). This approach was taken in
Refs.~\refcite{Komar_1959,Arnowitt_1961,Bondi_1962,
Sachs_1962,Newman_1962,Janis_1965,Geroch_1970,Hansen_1974,
Thorne_1980,Simon_1983}, for reviews see Ref.~\refcite{Thorne_1980,
Quevedo_1990}. The second type of definitions characterizes solely the source
and will be called \emph{source multipole moments} or \emph{source integrals}.
Such definitions were put forward, e.g. in
Ref.~\refcite{Dixon_1973,Ashtekar_2004}.

In this paper, we show that both types of definitions can be related to each
other also in general relativity, i.e., we give a source integral formulation of
asymptotic multipole moments. Details of the calculations can be found in Refs.
\refcite{Gurlebeck_2012,Gurlebeck_2013a}. Such a formalism has many
applications, e.g. in the description of tidal deformations of neutron stars
and black holes. It also allows to characterize exact and
numerical solutions by their asymptotic multipole moments using only the metric in the
interior of the matter. We give an example of that in the end of this paper.
 
\section{Formalism}

 We concentrate on axially symmetric and static spacetimes. In
 Weyl form they read:
 \begin{align*}
   ds^2=\mathrm{e}^{2 k-2U}\left(d\rho^2+d\zeta^2\right)+
   W^2\mathrm{e}^{-2U} d\varphi^2-\mathrm{e}^{2U}d t^2.
 \end{align*}
The metric functions $U,~W$ and $k$ depend only on the Weyl coordinates $\rho$ and $\zeta$.
With the time-like and hypersurface orthogonal Killing vector $\xi^a$ and
the space-like Killing vector $\eta^a$, the metric functions are given by
\begin{align*}
  \e{2U}=-\xi_a\xi^a,\quad W^2=-\xi_a\xi^a\eta_b\eta^b.
\end{align*}
Furthermore, the Killing vectors define a 1-form
 \begin{align*}
 Z_a=\epsilon_{abcd}W^{,b}W^{-1}\eta^c\xi^d
 \end{align*}
that is hypersurface orthogonal everywhere as well as exact in the vacuum
region. Hence, a potential $Z$ and an integrating factor $X$ exist such
that $Z_{,a}=X Z_{a}$, where $X=1$ in the exterior of a topological
2-sphere $\mathcal S_0$, which encloses all sources. In the vacuum region
and in \emph{canonical} Weyl coordinates, we have
 $Z=\zeta+\mathrm{const.}$ and $W=\rho$. Since we can shift the
 $\zeta$-coordinate freely, we can drop the constant of integration. It
 specifies the origin with respect to which the asymptotic
 multipole moments are introduced. Using the canonical Weyl coordinates, we can define
 Weyl's asymptotic multipole moments $U^{(r)}$ covariantly via an expansion of $U$ along
 the axis of symmetry $\rho=0$ close to spatial infinity:
 \begin{align*}
 U(\rho=0,\zeta)=\sum\limits^{\infty}_{r=0}U^{(r)}|\zeta|^{-r-1}.
 \end{align*}
The $U^{(r)}$ are sufficient to calculate the Geroch-Hansen multipole
moments\cite{Fodor_1989}.
 
Let us further introduce two sets of polynomials in $W$ and $Z$:
\begin{align*}
\begin{split}
   N_{r}^{-}&=
   \sum\limits_{k=0}^{\left\lfloor\frac{r}{2}\right\rfloor}\frac{2(-1)^{k+1}
   r!W^{2k+1}Z^{r-2k}}{4^k (k!)^2(r-2k)!},~
   N_{r}^{+}=
   \sum\limits_{k=0}^{\left\lfloor\frac{r-1}{2}\right\rfloor}\frac{2
   (-1)^{k+1}r!W^{2k+2}Z^{r-2k-1}}{4^k (k!)^2(r-2k-1)!(2k+2)}.
\end{split}
\end{align*}
$\lfloor x \rfloor$ denotes the greatest integer $y$ with $y\leq x$. Moreover, suppose that each black hole is enclosed by a topological 2-sphere $\mathcal S_i$ that contains no other sources and assume that there is a (not necessarily connected) region $\mathcal V$ in the projection orthogonal to $\xi^a$ that covers all matter. Then the source integrals evaluate to\cite{Gurlebeck_2012,Gurlebeck_2013a}  
\begin{align}\label{eq:source_integrals}
\begin{split}
   U_{r}=&\frac{1}{8\pi}\int\limits_{\mathcal V}\rho_r \mathrm d\mathcal V+\frac{1}{8\pi}\sum\limits_i \int\limits_{{\mathcal   S}_i} \frac{\e{U}}{W}\left(N_{r}^{-}U_{,\hat n}-N^{+}_{r,W}Z_{,\hat
     n}U+N^{+}_{r,Z}W_{,\hat n}U\right) \mathrm d\mathcal S_i,\\
   \rho_r=&\e{U}\left[-\frac{N^{-}_{r}}{W}
     R_{ab}\frac{\xi^a\xi^b}{\xi^c\xi_c} +
     N^{+}_{r,Z}U  \left(\frac{W^{,a}}{W\phantom{{}^a}}
     \right)_{;a} - N^{+}_{r,W}U\left(\frac{Z^{,a}}
     {W\phantom{{}^a}} \right)_{;a} +\right.\\
  &\left.N^{+}_{r,WZ}\frac{U}{W}
     \big(W^{,a}W_{,a}- Z^{,a}Z_{,a}\big)\right],
\end{split}
\end{align}
where $\mathrm d\mathcal S_i$ and $\mathrm d\mathcal V$ denote the proper surface or volume element of $\mathcal S_i$ and $\mathcal V$, respectively. $f_{,\hat n}$ is the derivative of $f$ in the direction of the unit normal of $\mathcal S_i$. 

 We conclude the paper discussing one possible application of the source
 integrals \eqref{eq:source_integrals}. Assume a matter distribution is given
 and the metric is known in $\mathcal V$. Even then it is far from trivial (at
 least in the stationary case\cite{Ansorg_2002}) to obtain a global
 asymptotically flat solution, if it exists. The source integrals provide a tool
 to solve this task. We demonstrated the procedure for static dust
 configurations in Refs.~\refcite{Gurlebeck_2012,Gurlebeck_2013a}. There we
 showed that all asymptotic multipole moments of static and axially symmetric
 dust configurations can be calculated and that they all vanish. This
 contradicts the presence of a dust distribution with a positive mass density.
 Of course, this result is already known and more general non-existence results
 for dust including the rotating case can be found in
 Refs.~\refcite{Gurlebeck_2009,Pfister_2010} and references therein. Although
 the non-existence was proved\cite{Gurlebeck_2012,Gurlebeck_2013a}, this example
 shows in a concise way how the source integrals can be applied in more
 difficult physical situations like rotating relativistic stars.
 This and other applications, e.g. to tidal distortions of black holes, will be
 investigated in future work.

\section*{Acknowledgment} The author gratefully acknowledges support from the
DFG within the Research Training Group 1620 ``Models of Gravity'' and from the Grant
GA\v CR-202/09/0772. The author thanks C. L\"ammerzahl, V. Perlick and O. Sv\'itek for helpful discussions.


\end{document}